\documentclass[aps,pra,epsfigure,twocolumn,showpacs]{revtex4}
\usepackage{dcolumn}    
\usepackage{bm} 
\usepackage{graphicx}
\usepackage{amsmath}    
\usepackage{latexsym}
\usepackage{amsfonts}   
\usepackage{amssymb}
\usepackage{array}      
\usepackage{epsfig}
\usepackage{epstopdf}

\newcommand{\ket}[1]{\left\vert#1\right\rangle}
\newcommand{\miniket}[1]{\vert#1\rangle}

\newcommand{\id}{\;\mbox{$\rm{I} \hspace{-2.0mm} {\bf I}$}\,}
\newcommand{\miniid}{\;\mbox{$\rm{I} \hspace{-1.5mm} {\bf I}$}\,}

\newcommand{\minisand}[3]{\langle#1\vert#2\vert#3\rangle}

\begin{document}

\title{Quantum state transfer via temporal kicking of information}
%\title{Bang-bang kicking quantum state transfer}

\author{C. Di Franco$^1$, M. Paternostro$^2$, and M. S. Kim$^2$}

\affiliation{$^1$Department of Physics, University College Cork, Cork, Republic of Ireland\\
$^2$School of Mathematics and Physics, Queen's University, Belfast BT7 1NN, United Kingdom}

\begin{abstract}
We propose a strategy for perfect state transfer in spin chains based on the use of an unmodulated coupling Hamiltonian whose coefficients are explicitly time dependent. We show that, if specific and non-demanding conditions are satisfied by the temporal behavior of the coupling strengths, our model allows perfect state transfer. The paradigma put forward by our proposal holds the promises to set an alternative standard to the use of clever encoding and coupling-strength engineering for perfect state transfer.
\end{abstract}

\date{\today}
 \maketitle

Multiple-spin systems, in particular spin chains, have recently been the object of extensive studies. From the quantum information processing (QIP) viewpoint, such systems embody valuable media for quantum protocols. In fact, it has been found that specific forms of built-in and permanent intra-register couplings, such as those typical of spin-chain models, could be used for the purposes of quantum computation~\cite{alwayson} and communication in quantum networks. This second possibility is particularly interesting. In fact, while photons are ideal candidates for the long-haul transmission of information among different local nodes~\cite{flying}, their use in hybrid architectures for quantum networks (where the nodes are embodied by matter-like systems) requires the use of quantum interfaces. Despite the impressive experimental successes reported in this area, the realization of an interface is usually accompanied by technical problems and errors in transmission. On the other hand, for short-distance quantum communication, an alternative arrangement where information carriers and processors are embodied by physical systems having the same nature could be more advantageous. In the seminal work in Ref.~\cite{unmodulated}, the idea of using spin chains as {\it quantum wires} has been proposed as a way to avoid the interfacing problems mentioned above. The original idea has then been extended along various directions (for more details, see Ref.~\cite{bosereview}). In particular, Refs.~\cite{cambridge} showed that, by properly engineering the strength of the couplings in the chain, perfect state transfer could be achieved. A proposal to bypass the initialization of the medium by means of local operations and measurements on the extremal spins of the chain has been recently put forward~\cite{roastedchicken}. A similar result can be obtained by properly encoding the state to transmit in two spins at one end of the chain~\cite{marcin}.

Here, an alternative strategy to the one proposed in Ref.~\cite{cambridge} is presented. Instead of the pre-arrangement of the interaction strengths across the chain, we consider a uniform distribution of time-dependent couplings. We show that perfect transfer of information is achieved in our time-dependent architecture as well and that a formal mapping of the dynamics achieved in the two models is in order, thus letting our proposal emerge as an alternative paradigma for perfect quantum state transfer. We also study a scenario where the coupling strengths of the Hamiltonian are constant, while an external time-dependent magnetic field is used in order to provide the necessary temporal modulation required for information transmission. We believe that the introduction of a time-dependent term in the intra-chain couplings allows to bypass the intrinsic rigidity of protocols based on unmodulated and pre-engineered spin media. In fact, it is usually the case that spin-chain models are arranged so as to implement a specific communication or computational task and cannot be ``recycled" and used for a different one. Moreover, should an experimentally-prepared pattern of coupling strengths be found not accurate enough, the whole medium should be discarded (unless a non-ideal performance of the quantum protocol could be tolerated). Differently, by allowing the presence of time-dependent Hamiltonian terms, a more dynamical ``adjustment" of the medium performances would be possible by implementing a simple feedback loop: state transmission can be tested using a given functional form of the time-dependent terms of the Hamiltonian and, according to the results, this can be tuned so as to converge toward better performances. A time-dependent scheme for quantum state transfer has been discussed by Lyakhov and Bruder in Ref.~\cite{bruder}. Building up on general result described in Ref.~\cite{polacchi}, they only let the first and last spin-pair of a chain to experience time-dependent couplings. Our scheme, on the other hand, is different as we either consider a full set of time-dependent interaction strengths or a fully homogeneous pattern of constant couplings with a time-dependent external magnetic field. 

The remainder of this article is organized as follows. In Sec.~\ref{system} we describe the model used in our proposal. In Sec.~\ref{informationflux} we shortly review the main features of the information-flux approach, which is the main tool used in our investigation, and adapt it to our time-dependent analysis. Sec.~\ref{results} illustrates the way we obtain perfect state transfer when the coupling strengths depend on the interaction time via sharply rising/lowering pulses, while Sec.~\ref{realistic} is dedicated to the investigation of a much more realistic setting that considers finite rising times and pulse duration. In Sec.~\ref{multipartite}, the proposed model is exploited in order to create a genuine multi-partite entangled state. Finally, Sec.~\ref{remarks} summarizes our results.

\section{The model}
\label{system}

The system we analyze is an open spin-chain of $N$ elements, whose Hamiltonian reads
\begin{equation}
\label{modelloXY}
\hat{{\cal
H}}=\sum^{N-1}_{i=1}[J_x(t)\hat{X}_{i}\hat{X}_{i+1}+J_y(t)\hat{Y}_{i}\hat{Y}_{i+1}]+\sum^N_{i=1}B(t)\hat{Z}_i.
\end{equation}
Here, $J_x(t)$ and $J_y(t)$ are the coupling strengths of the pairwise interaction between adjacent spins and $B(t)$ is a magnetic field. In our notation, $\hat{X},\,\hat{Y}$ and $\hat{Z}$ denote respectively the $x,\,y$ and $z$-Pauli matrix. For the sake of simplicity, we consider $N$ as an odd number. However, all the results presented here can be straightforwardly adapted to the case of an even number of spins in the chain. Physical units are chosen throughout the article so that ${\hbar}=1$. It is important to note that the inter-spin couplings and the amplitude of the magnetic field are site independent. This is a feature that differenciates the system at hand from a few previous proposals for perfect state transfer available in the literature~\cite{cambridge,dimitris,roastedchicken,marcin}. Although Eq.~(\ref{modelloXY}) is spatially unmodulated (in analogy with~\cite{unmodulated}), the price to pay in order to achieve unit transfer fidelity for any length of the chain is the time dependence of the interaction strengths. We will show later in this article that such a request can also be relaxed: perfect state transfer can be achieved even if only the magnetic field is time-dependent while any other term is constant.

Eq.~(\ref{modelloXY}) does not preserve the number of spin-excitations as it does not commute with the total $z$-component of the spin of the system. Frequently, when one faces the case of dynamics ruled by spin-preserving models, on the assumption of proper boundary conditions, it is convenient to diagonalize the coupling Hamiltonian by means of a sequence of operation comprising Wigner-Jordan, Fourier, and Bogoliubov transforms~\cite{lieb}. In our case, due to the spin-non preserving nature of $\hat{\cal H}$, the mutual coupling among sectors of the Hilbert space labeled by different quantum spin numbers has to be considered. Rather than applying techniques for the exact diagonalization of Eq.~(\ref{modelloXY}), here we tackle the evolution of the system by means of an {\it information-flux approach}, which is specifically designed for multi-spin interactions~\cite{informationflux,informationfluxijqi}. Our method does not rely on the explicit analysis of the energy spectrum of the chain and allows us to gather an intuitive picture of the dynamics at hand. Such an approach has already proved its flexibility and efficiency in identifying Hamiltonian configurations suitable for state transfer and entanglement generation~\cite{dimitris,matryoshka,roastedchicken}.

\section{Information-flux approach}
\label{informationflux}

The information-flux approach requires the time-evolved form of specific operators $\hat{\cal O}_j$ in the Heisenberg picture, which is given by
\begin{equation}
\hat{\cal O}_j(t)=\hat{\cal U}(t)^\dag\hat{O}_j\,\hat{\cal U}(t),
\end{equation}
with $\hat{\cal U}(t)$ the time evolution operator, $\hat{O}=\hat{X},\hat{Y},\hat{Z}$ and $j=1,...,N$. Using this method, one is able to understand the dependence of $\hat{\cal O}_j(t)$ on any $\hat{O}_k$ and design the proper set of interactions in such a way that it becomes possible to {\it drive} a desired evolution by means of engineered quantum interference~\cite{informationflux}. In our case, we will exploit the time dependence of the couplings in order to induce precisely this effect.

Our task here is the transmission of quantum information from the first spin to the last one in the chain. This requires the study of $\hat{\cal O}_N(t)$'s, which can be decomposed into the operator basis built out of all possible tensor products of $\{\hat{\id}_i,\hat{X}_i,\hat{Y}_i,\hat{Z}_i\}$ with $\hat{\id}_i$ the $2 \times 2$ identity operator applied to spin $i$. We have
\begin{equation}
\minisand{\Psi_0}{\hat{\cal O}_N(t)}{\Psi_0}\!=\!\sum_{O'=X,Y,Z,{\scriptsize\miniid}}{\cal I}^{OO'}(t)\minisand{\phi_0}{\hat{O}'_1}{\phi_0},
\end{equation}
where $\ket{\Psi_0}=\ket{\phi_0}_1\otimes\ket{\psi_0}_{2...N}$ is the initial state of the whole chain, the first spin being in $\ket{\phi_0}_1$ and the rest of the chain in $\ket{\psi_0}_{2...N}$ (this definition can be easily generalized to the case of mixed states of the chain). The coefficient ${\cal I}^{OO'}(t)$ is defined as the information flux at time $t$ from $\hat{O}'_1$ to $\hat{O}_N$~\cite{informationflux}. If our system achieves $|{\cal I}^{OO'}|=1$ when $O'=O$ and for $O=X,Y,Z$, we have perfect $1\rightarrow{N}$ state transfer.

\begin{figure}[t]
\centerline{\psfig{figure=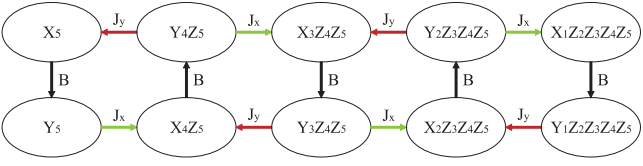,width=8.5cm}}
\caption{Oriented graph describing how the operators $\hat{X}_5$ and $\hat{Y}_5$ evolve in the Heisenberg picture. The operators within each circle (a node) give rise to their nearest neighbors on commutation with $\hat{{\cal H}}$ in Eq.~(\ref{modelloXY}). The oriented edges connect such nodes. The corresponding coefficients are also shown and an outgoing (incoming) edge with respect to a node implies a $+$ $(-)$ sign. Black, light green and dark red edges are associated respectively with $B$, $J_x$ and $J_y$ interaction terms.}
\label{doublechain}
\end{figure}

We can determine which are the terms of the operator basis that are involved in the decomposition of $\hat{\cal O}_N(t)$'s by representing such operators in an oriented weighted graph, whose construction is straightforward. Each node of the graph corresponds to an operator involved in the decomposition. If the  interest is focused on the evolution of $\hat{O}_N$, it should occupy the first node of the graph. Any operator resulting from the commutators of $\hat{O}_N$ with the Hamiltonian should be included in the graph and linked to it with an oriented edge. These steps are repeated as necessary and the construction process ends when no new operator is created on commutation. A numerical weight, determined by the performed commutations and needed in order to calculate the information flux, is attached to each edge of the graph. Finally, an outgoing (incoming) edge corresponds to a + (-) sign. Due to the particular form of $\hat{\cal{H}}$ in Eq.~(\ref{modelloXY}), $\hat{\cal X}_N$ and $\hat{\cal Y}_N$ will generate the oriented graph shown in Fig.~\ref{doublechain} (where the case $N=5$ has been considered). Clearly, $\hat{Y}_N$ already appears in the graph corresponding to $\hat{X}_N$ and we do not need to construct a dedicated graph for it.

For the sake of clarity, here we show how to evaluate the information flux for a time-independent Hamiltonian such as Eq.~(\ref{modelloXY}) for $J_{x,y}(t)=J_{x,y}$. The method can be explicitly extended to any time-dependent model. However, in our numerical analysis we divide the total time of the evolution in a series of short steps within which the Hamiltonian is taken as constant. In this way, in the limit of a large number of steps and correspondingly short time steps, the flux of information can be accurately evaluated by means of only a light computational effort. A cut-off in the number of considered temporal steps is chosen so as to have no noticeable difference in the results if a larger number is taken. For a time-independent Hamiltonian, $\hat{\cal O}_j(t)=e^{i\hat{\cal H}t}\,\hat{O}_j\,e^{-i\hat{\cal H}t}$ and, by means of the operator expansion formula~\cite{mandelwolf}, we have
\begin{equation}
\hat{\cal O}_j(t)=\sum^\infty_{k=0}\frac{(it)^k}{k!}\hat{\cal C}_k(\hat{O}_j)
\end{equation}
with $\hat{\cal C}_k(\hat{O}_j)$ the nested commutator of order $k$ between $\hat{\cal H}$ and $\hat{O}_j$ and $\hat{\cal C}_0(\hat{O}_j)=\hat{O}_j$.

Here, for the sake of a clear presentation of our proposal and without loss of generality, we focus on the case of $N=5$ (the case of an arbitrary $N$ being easily extrapolated from our example) and study the dynamics of  $\hat{X}_5$ under the model in Eq.~(\ref{modelloXY}). The terms in the operator expansion are deduced by 
\begin{equation}
 \begin{split}
\hat{\cal C}_1(\hat{X}_5)&= [{\hat{\cal H}},\hat{X}_5]=-2i\,J_y\,\hat{Y}_4\hat{Z}_5+2i\,B\,\hat{Y}_5,\\
 \hat{\cal C}_2(\hat{X}_5)&=[{\hat{\cal H}},[{\hat{\cal H}},\hat{X}_5]]=2iB(2i\,J_x\,\hat{X}_4\hat{Z}_5-2i\,B\,\hat{X}_5)\\
&-2i\,J_y(2i\,J_y\,\hat{X}_5+2i\,J_x\,\hat{X}_3\hat{Z}_4\hat{Z}_5-2i\,B\,\hat{X}_4\hat{Z}_5),\\
  &\vdots
 \end{split}
\label{commutators}
\end{equation}
Clearly, the only operators involved in this iterative sequence are those in the graph of Fig.~\ref{doublechain}. Therefore, it is possible to write the evolved operators $\hat{\cal X}_5(t)$ as
\begin{equation}
\label{evoluti}
 \begin{split}
  \hat{\cal X}_5(t)=&\alpha_1(t)\hat{X}_5+\alpha_2(t)\hat{Y}_4\hat{Z}_5+\alpha_3(t)\hat{X}_3\hat{Z}_4\hat{Z}_5+\\
  &\alpha_4(t)\hat{Y}_2\hat{Z}_3\hat{Z}_4\hat{Z}_5+\alpha_5(t) \hat{X}_1\hat{Z}_2\hat{Z}_3\hat{Z}_4\hat{Z}_5+\\
  &\alpha_6(t)\hat{Y}_5+\alpha_7(t)\hat{X}_4\hat{Z}_5+\alpha_8(t)\hat{Y}_3\hat{Z}_4\hat{Z}_5+\\
  &\alpha_9(t)\hat{X}_2\hat{Z}_3\hat{Z}_4\hat{Z}_5+\alpha_{10}(t)\hat{Y}_1\hat{Z}_2\hat{Z}_3\hat{Z}_4\hat{Z}_5.
 \end{split}
\end{equation}
When the parameters $\alpha_j(t)$ cannot be analytically evaluated due to the difficulties of the evolution, it is still possible to approximate them by means of recurrence formulas. We have
\begin{equation}
\alpha_j(t)\sim\sum_{l=0}^M\frac{(2t)^l}{l!}\alpha_j^{(l)},
\end{equation}
where $M$ is a proper cut-off and
\begin{equation}
 \begin{split}
\alpha_1^{(l)}&=-J_y\alpha_2^{(l-1)}+B\alpha_6^{(l-1)},\\
\alpha_2^{(l)}&=J_y\alpha_1^{(l-1)}+J_x\alpha_3^{(l-1)}-B\alpha_7^{(l-1)},\\
  &\vdots
 \end{split}
\end{equation}
with $\alpha_j^{(0)}=0$ ($1$) for $j\ne1$ ($j=1$). It is immediate to note that these recurrence formulas come directly from the commutation rules in Eq.~(\ref{commutators}). They can thus be easily derived from the graph in Fig.~\ref{doublechain}. The same formulas, but with the initial conditions $\alpha_j^{(0)}=0$ ($1$) for $j\ne6$ ($j=6$), describe the evolution of the operator $\hat{Y}_5$. The approach described so far constitutes the basis for the analysis of perfect quantum state transfer via temporally controlled information kicking, which is the focus of the next Section.

\section{The ideal case}
\label{results}

\begin{figure}[b]
\centerline{\psfig{figure=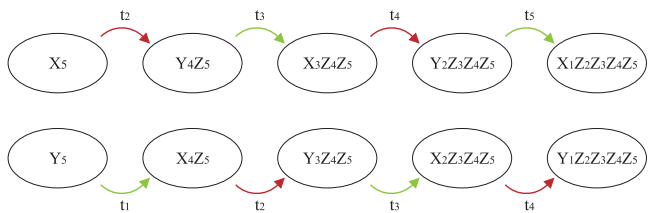,width=8.5cm}}
\caption{``Operator jumps'' induced by ${\hat{\cal H}}$ in Eq.~(\ref{modelloXY}), when $J_x$ and $J_y$ are alternately different from zero. $J_x$ ($J_y$) is non-zero in the time intervals $t_i$ with an odd (even) value of $i$, and its integral over each of these intervals is equal to $\pi/4$.}
\label{doublechainjumps}
\end{figure}

Let us study the action of ${\hat{\cal H}}$ in Eq.~(\ref{modelloXY}) when only one of the inter-spin  coupling terms differs from zero. The graph in Fig.~\ref{doublechain} shows that, in these conditions, every node is linked to just one nearest neighbor. In fact, the full oriented graph for the information flux analysis corresponding to Eq.~(\ref{modelloXY}) can be seen as the juxtaposition of three mutually disconnected subgraphs, each associated with only one of the coupling terms in $\hat{\cal H}$ and represented by a different color in Fig.~\ref{doublechain}. It is easy to track the operator dynamics associated with these configurations. Let us consider, for instance, the case $B(t)\neq{0}$ and $J_{x,y}(t)=0~\forall{t}$. Fig.~\ref{doublechain} (black color) shows that $\hat{X}_5$ and $\hat{Y}_5$ are mutually linked and, by using the formal quantitative analysis of the previous Section, we have
\begin{equation}
\begin{split}
&\hat{\cal X}_5(t)=\cos[2\beta(t)]\hat{X}_5+\sin[2\beta(t)]\hat{Y}_5,\\
&\hat{\cal Y}_5(t)=-\sin[2\beta(t)]\hat{X}_5+\cos[2\beta(t)]\hat{Y}_5
\end{split}
\end{equation}
with $\beta(t)=\int_0^t{B}(t')dt'$. Clearly, an isolated node in a graph corresponds to an operator that is constant in time [for instance, $\hat{\cal X}_5(t)=\hat{X}_5$ when $J_{x}\ne0$, $J_{y}=B=0$, as in Fig.~\ref{doublechain} (light green)]. This result is key to our investigation. If we let the interaction terms be alternatively non-zero for a time window $[0,\tau]$ such that $\beta(\tau)=\pi/4$, the action of ${\hat{\cal H}}$ in Eq.~(\ref{modelloXY}) will correspond to {\it kicks of information} between pairs of connected nodes of the graphs discussed above. In Fig.~\ref{doublechainjumps} we show an example for $J_{x,y}(t)$ being alternately non-zero for five time intervals such that $\int^{t_{j}}_{t_{j-1}}J_{x}(t')dt'=\pi/4$ for $j=1,3,5$ [$\int^{t_{j}}_{t_{j-1}}J_{y}(t')dt'=\pi/4$ for $j=2,4$], with $J_x(t)=0$ [$J_y(t)=0$] whenever $J_y(t)\ne0$ [$J_x(t)\ne0$]. The negligibility of the mutual overlaps of the coupling functions ensures that the information kicks occur according to the ordered scheme in Fig.~\ref{doublechainjumps}.
A similar transport of information from the leftmost to the rightmost part of the graph is achieved by kicking information using a pattern given by fixing $J_y=0$ ($J_x=0$) and alternately changing the amplitude of $J_x$ ($J_y$) and $B$ in a way so as to satisfy the conditions stated above. After a sequence of $N$ (for time-dependent $J_x$ and $J_y$) or $2N-1$ kicks (for time-dependent $B$ and either $J_x$ or $J_y$), one obtains
\begin{equation}
\begin{split}
&\hat{\cal X}_5(\tau^*)=X_1Z_2Z_3Z_4Z_5,\\
&\hat{\cal Y}_5(\tau^*)=Y_1Z_2Z_3Z_4Z_5,
\end{split}
\label{extremalspinevolution}
\end{equation}
where $\tau^*$ is the total time of the evolution. By initially preparing the $j^{\text{th}}$ spin of the chain (except the first) in an eigenstate of $\hat{Z}_j$, there is a complete end-to-end transport of information across the chain. This evolution interaction can thus be exploited for a perfect state transfer exactly as it happens for the Hamiltonian model of Ref.~\cite{cambridge}. Quite interestingly, any other state-transfer protocol relying on the same information-flux structure, as those in Refs.~\cite{roastedchicken,marcin}, can be mapped into the very same temporal pattern of information kicking illustrated here. This is the main result of our investigation, proving that an alternative to the standard architecture of properly arranged interaction-strength patterns for quantum state transfer is actually possible. The key is the introduction of an alternated series of non-overlapping temporal kicks of information. The equivalence of the two scenarios has here been proven in terms of an isomorphism of the respective information fluxes and is, as such, completely general. Depending on the details of a specific practical realization of the intra-chain couplings, temporal control of the form highlighted here can be more convenient than pre-fabrication of a specific pattern of static coupling strengths. This is the case, for instance, for quasi-unidimensional optical lattices loaded  with neutral atoms, where inter-site couplings are achievable in a time-controlled way via external optical potentials or cold atomic collisions. In perspective, this could also be a viable option in solid-state structures (such as arrays of Josephson junctions), where fabrication of exactly the pattern required for ideal state transfer would be demanding (if not prohibitive) already at moderate chain lengths. Achieving control via proper voltage/magnetic pulses inducing time-controlled information kicks would be a possibility to exploit instead.

\section{The realistic case}
\label{realistic}
In order to go beyond the idealization of the scheme assessed in the previous Section and make our analysis closer to more realistic situations, the ideal conditions on the pattern of time-dependent couplings invoked before will be relaxed here. Let us first consider $J_y(t)=0~\forall{t}$ with $J_x(t)$ and $B(t)$ following the behavior dictated by 
\begin{equation}
\begin{split}
&J_x(t)=J_{max}[\sin(t+\pi/4)]^m,\\
&B(t)=B_{max}[\cos(t+\pi/4)]^m,
\end{split}
\label{sincos}
\end{equation}
where $J_{max}=B_{max}$ are the maximum values of $J_x(t)$ and $B(t)$ (properly chosen in order to satisfy the integral condition) and $m$ is an even number. Pulse shapes of this form (for moderate values of $m$) are routinely generated by commercial pulse generators. A large value of $m$ obviously implies small overlap between the two coupling functions.  A period of $2\pi$ will correspond to a ``horizontal'' and a ``vertical'' kick in the associated operator graph. For a numerical analysis of the information-flux behavior under the action of this time-dependent Hamiltonian, we divided the total time interval $[0,2N\pi]$ in short steps. Within each of them, both $J_x(t)$ and $B(t)$ are taken as constant and numerically equal to the average values they assume in the respective time window. In this way, the recurrence formulas presented in Sec.~\ref{informationflux} allow us to estimate the evolution of $\hat{X}_N$ or $\hat{Y}_N$ for the duration of the step. Clearly, the shorter is the time step, the more accurate is the estimate. The method is then iterated for each step and the total evolution of $\hat{X}_N$ or $\hat{Y}_N$ is finally reconstructed. As the number of operators involved in each decomposition is equal to $2N$, a light computational power is required.

\begin{figure}[t]
\centerline{{\bf (a)}\hskip4cm{\bf (b)}}
\centerline{\psfig{figure=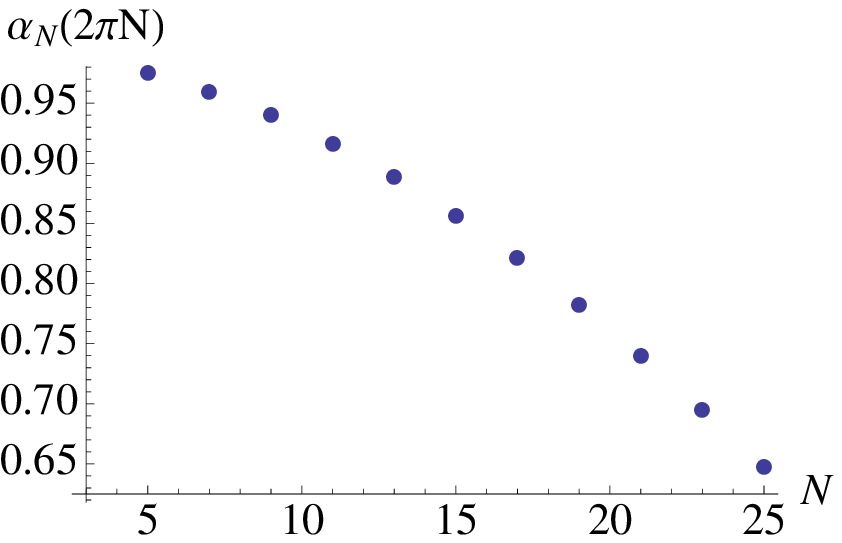,width=4cm}\hskip0.5cm\psfig{figure=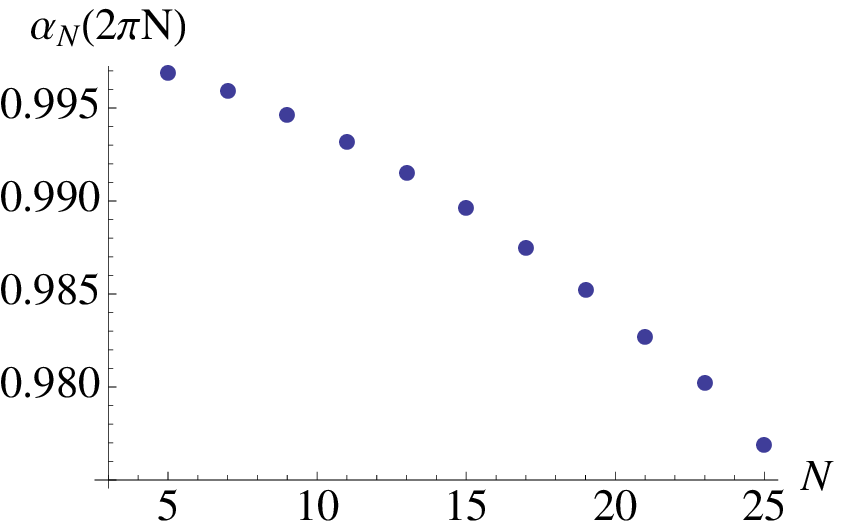,width=4cm}}
\caption{Maximum value of $\alpha_N(t)$ against the number of spins $N$ in the chain. The value of $m$ in Eq.~(\ref{sincos}) is respectively equal to 4 [panel {\bf(a)}] and 6 [panel {\bf(b)}].}
\label{plotpowers}
\end{figure}

In order to evaluate the performance of our scheme, both the coefficient $\alpha_{N}(t)$, attached to $\hat{X}_1\hat{Z}_2\cdot\cdot\cdot\hat{Z}_N$ in the decomposition of $\hat{\cal X}_N(t)$, and $\alpha_{2N}(t)$, attached to $\hat{Y}_1\hat{Z}_2\cdot\cdot\cdot\hat{Z}_N$ in the decomposition of $\hat{\cal Y}_N(t)$, should be studied. However, one can equally well characterize the whole state transfer process simply by considering $\alpha_N(t)$ only. The maximum value of $\alpha_N(t)$ for two values of $m$ is studied against $N$ in Figs.~\ref{plotpowers} {\bf(a)} and {\bf(b)}.
It is matter of performing an explicit calculation to see that the maximum of $\alpha_{2N}(t)$ follows similar behaviors. We then quantify the mean transmission fidelity $\overline{F}(N,t)$ by averaging the state fidelity $\langle{\psi_{\text{in}}}|\rho_N(t)|{\psi_{\text{in}}}\rangle$ between a pure input state $\ket{\psi_{\text{in}}}$ to transfer and the state $\rho_N(t)$ of spin $N$ at time $t$. By assuming a uniform distribution of $|\psi_{\text{in}}\rangle$'s~\cite{unmodulated}, we find 
\begin{equation}
\overline{F}(N,2N\pi)=\frac{1}{2}\left[1+\alpha_N(2N\pi)\left(\frac{2}{3}+\frac{1}{3}\alpha_N(2N\pi)\right)\right].
\end{equation}
For instance, if we take $m=6$, we obtain for any length of the chain $N\le25$ a value of $\overline{F}(N,2N\pi)>0.984$. As $\alpha_{N}(2N\pi)\rightarrow{1}$, the average fidelity converges to $1$, resulting in perfect state transfer.

\begin{figure}[b]
\centerline{\psfig{figure=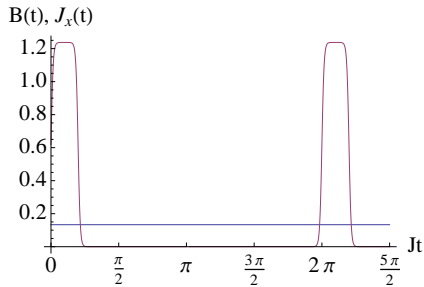,width=5.5cm}}
\caption{Temporal behavior of $B$ (red line) and $J_x$ (blue line), for a value of $\delta=5$. The quantities plotted in the vertical axis are in units of $J$.}
\label{plotbandjbehavior}
\end{figure}

\begin{figure}[t]
\vskip0.25cm
\centerline{\psfig{figure=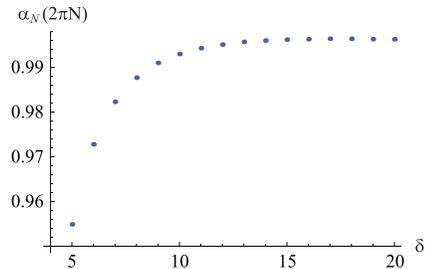,width=5.5cm}}
\caption{Maximum value of $\alpha_N(t)$ against the parameter $\delta$, for a chain with $N=5$ spins.}
\label{5qubitsdelta5to20}
\end{figure}

\begin{figure}[t]
\vskip0.25cm
\centerline{\psfig{figure=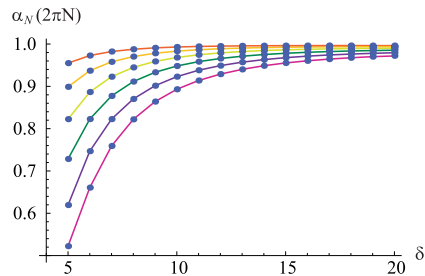,width=5.5cm}}
\caption{Maximum of $\alpha_N(t)$ against the parameter $\delta$, for $N=2k+1$ with $k=2\rightarrow7$ (going from the top to the bottom curve).}
\label{5to15qubitsdelta5to20}
\end{figure}

The ability to temporally control both $J_x$ and $B$ can however be experimentally demanding. In particular, while the arrangement of external potentials with a desired time-dependence can be relatively straightforward, this might not be the case for $J_{x}(t)$'s, which are in general determined by coupling mechanisms internal to the chain itself and hardly controllable with the degree of accuracy required by our scheme. We thus make our requests lighter by taking $J_x$ as constant and allowing a time dependence only for the external magnetic field. Obviously, it is formally the same to take a constant $B$ with time-varying $J_x(t)$. However, pragmatically our choice seems to us to be the least demanding. We thus adapt the strategy assessed before to the case of $J_x=J_{\text{const}}$. Our quantitative study shows that a slight modification to the sinusoidal trend described before is sufficient to retain the efficiency of the protocol. We take well-spaced square-shaped pulses for $B(t)$ as shown in Fig.~\ref{plotbandjbehavior}.
The parameter $\delta$ quantifies the width of magnetic field pulses. It can be roughly seen as half of the overall period divided by the time in which $B$ is close to its maximum value. Clearly, a large value of $\delta$ corresponds to a more ideal case, as the overlap between the two functions is smaller, hence affecting the maximum value that $\alpha_N(t)$ can achieve and, in turn, the transfer fidelity.  The dependence of $\alpha_{N}(2N\pi)$ on $\delta$ is shown in Fig.~\ref{5qubitsdelta5to20} for $N=5$.
The fidelity $\overline{F}(5,10\pi)$ is already larger than $0.99$ for $\delta=8$. Of course, the effect of the overlap between the two functions depends on $N$: the longer the chain, the lower the transmission fidelity. The dependence of the fidelity on the length of the chain is plotted in Fig.~\ref{5to15qubitsdelta5to20}.
We note that, for $N$ up to $15$, a value of $\delta=16$ is large enough to obtain the maximum of $\alpha_N$ $\sim0.96$ (corresponding to $\overline{F}\sim0.974$). 
 
\begin{figure}[b]
\centerline{\psfig{figure=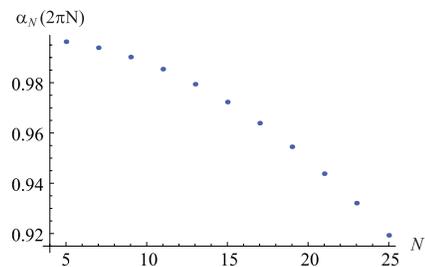,width=5.5cm}}
\caption{Maximum of $\alpha_N(t)$ against the length of the chain $N$ for $\delta=20$.}
\label{5to25qubitsdelta20}
\end{figure}

To give a complete overview of the protocol, one can finally analyze, for a fixed $\delta$, the behavior of the maximum of $\alpha_N$ against $N$. The case of $\delta=20$ has been considered, and the results are shown in Fig.~\ref{5to25qubitsdelta20}.
In these conditions, the corresponding transmission fidelity for a chain with $N=25$ spins is $\sim0.947$.

We have also studied the case in which the non-zero interaction terms are $J_x$ and $J_y$, with $B=0$. The transmission fidelity is larger than the one obtained with non-zero $J_x$ ($J_y$) and $B$, as the number of kicks required is smaller and the effect of the overlap is thus reduced. However, we presented the results for non-zero $J_x$ ($J_y$) and $B$ as we consider the choice of a time-dependent magnetic field the least demanding from an experimental point of view.

\section{Generation of multi-partite entanglement}
\label{multipartite}

In this Section we discuss another interesting aspect of the proposed model, namely the possibility to generate genuine multi-partite entanglement. Bi-partite as well as multi-partite entanglement creation in spin chains has been recently studied from different points of view. For instance, one can generate entanglement by means of a proper pattern of coupling parameters~\cite{matryoshka}, by driving the evolution through a resonant interaction~\cite{david}, by performing a global quench~\cite{hannu}, or by leaving two empty sites in a uniformly filled chain~\cite{giulia}. By assuming the ability to initialize the state of the chain before the application of the pulse sequence in the protocol proposed in this article (we necessarily have more demanding requirements for this particular task), Greenberger-Horne-Zeilinger (GHZ) states can be generated~\cite{ghz}. Such a resource is well-known to be useful for multi-agent protocols for distributed QIP like quantum secret sharing, remote implementation of unknown operations and quantum average estimation~\cite{average}. Let us consider again the Hamiltonian $\hat{\cal H}$ in Eq.~(\ref{modelloXY}) under the ideal conditions mentioned above. It is interesting to notice that, after the proper kicking of information, the evolution of {\it all} the spin operators associated with the elements of the chain have exactly the same form as those obtained by using the model proposed in Refs.~\cite{cambridge,roastedchicken}.

We now want to evaluate the dynamics of a particular pure state of the chain under the action of this Hamiltonian. We will need to abandon the assumption that at most a single spin excitation populates the system. It is straightforward to check that, whenever spin $i$ is in an eigenstate of $\hat{Z}_i$, the corresponding symmetric spin $N-i+1$ will end up in the same state. Whenever two or more spins are initially in eigenstates of $\hat{X}$ or $\hat{Y}$, the corresponding symmetric spins will end up in an entangled state. Let us consider, for definiteness, an even number of spins. We call $\{a_1,a_2,...,a_n\}$ the indices corresponding to spins initially prepared in eigenstates $\ket{d}_{a_i}=\ket{0,1}_{a_i}$ of $\hat{Z}$ and $\{b_1,b_2,...,b_m\}$ those labeling spins initially prepared in eigenstates $\miniket{\tilde{d}}_{b_i}=\ket{+,-}_{b_i}$ of $\hat{X}$ (the analysis can be straightforwardly generalized to the case of odd number of spins and eigenstates of $\hat{Y}$). The initial state of the chain is thus
\begin{equation}
\ket{\Psi_0}=\bigotimes_{\{a_i,bi\}}\miniket{d,\tilde{d}}_{a_i,b_i}~~~(i=1,...,n).
\end{equation}
It is matter of a straightforward calculation to see that, in this case, the final state corresponds to a mirror-inversion operation on the state
\begin{equation}
\ket{\Psi(t)}=\frac{1}{\sqrt{2}}\bigotimes_{\{a_i\}}\ket{d}_{a_i}\otimes\left[\bigotimes_{\{b_i\}}\miniket{\tilde{d}}_{b_i}+i(-1)^l\bigotimes_{\{b_i\}}\hat{Z}_{b_i}\miniket{\tilde{d}}_{b_i}\right],
\end{equation}
with $l=0,1$ depending on the actual form of $\ket{\Psi_0}$. Clearly, $\ket{\Psi(t)}$ is the tensor product of a separable state for spins $\{a_1,a_2,...,a_n\}$ and an $m$-particle GHZ-like state for spins $\{b_1,b_2,...,b_m\}$.

\section{Remarks}
\label{remarks}

By means of the information flux approach, we have presented a new strategy for obtaining perfect state transfer in a finite, open chain of spins. We have shown that the evolution of spin operators under the action of a site-dependent Hamiltonian that is already known to allow perfect state transfer can be mimicked by the use of a homogeneous time-dependent Hamiltonian. The engineering of the coupling strengths, necessary in the first scenario, is thus converted to a temporal arrangement of the interaction terms. We have discussed a particularly interesting case where such a time dependence can be reduced to the requirement of just a non-constant magnetic field affecting the spins of the chain. Interestingly, this very same model can also be used in order to create a GHZ-like entangled state shared by a sub-set of spins of our choice, leaving all the other spins in a separable state. We believe that our proposal to achieve perfect state transfer and quantum correlation sharing in many-body registers of interacting spins widens the possibilities for efficient and non-demanding short/medium-haul quantum communication.

\acknowledgments 

We acknowledge support from the UK EPSRC and QIPIRC. C.D.F. is supported by the Irish Research Council for Science, Engineering and Technology. M.P. is supported by EPSRC (EP/G004579/1).

\end{document}